# Resolving inconsistencies of runtime configuration changes through change propagation and adjustments[1]

Azadeh Jahanbanifar[2], Ferhat Khendek[2], Maria Toeroe[3]

[2]Engineering and Computer Science, Concordia University
Montreal, Canada
{az_jahan, khendek}@encs.concordia.ca

[3]Ericsson Inc.
Montreal, Canada
maria.toeroe@ericsson.com

**Abstract:** A system configuration may be modified at runtime to adapt the system to changes in its environment or for fine-tuning. For instance, a system administrator may change a few entities/attributes in the configuration to improve error recovery and system availability. However, these changes focusing on fine-tuning availability may violate some global system constraints captured in the configuration and therefore break configuration consistency, system properties and operations. This is generally due to the partialness of the changes performed by the administrator who is not aware of all the entities/attributes being in relations with the modified entities/attributes. In this paper, we propose an approach for completing such partial sets of changes at runtime to resolve inconsistencies arising from those partial changes. This adjustment approach consists of the characterization of related entities/attributes and their modification to re-establish the configuration consistency. We achieve this by propagating the changes in the configuration according to the system constraints following the possible impacts of the configuration entities on each other. We aim at minimizing the complementary modifications to control the side-effects of the change propagation as we target the domain of highly available systems.

## 1. INTRODUCTION

A system configuration describes the system resources, their characteristics/attributes, and their relations that were defined to reflect certain properties, such as functionalities, performance, security, availability etc. The relations between the system's entities/attributes are captured in the configuration as system constraints, also referred to as consistency rules. They need to be maintained throughout the system life cycle despite any changes or reconfigurations, which, in case of high-availability systems or clouds systems, inevitably happen at runtime to adapt the system to the changes in its environment, for fine-tuning, and so on. In other words, the runtime changes should not violate the system properties captured in a set of constraints, also called consistency rules.

---

[1] This paper is an extension and a generalization of the work presented in [23].

Any reconfiguration request has to be checked against the system constraints. The requested changes are deemed unsafe and/or incomplete if any system constraint is violated. Validators can detect these violations and veto the proposed changes [1]. Violations due to incomplete changes may however be resolved with complementary modifications that re-establish configuration consistency. This is especially useful in the case of self-adaptive systems that need to adjust automatically to changes at runtime [10]. Nevertheless, finding the proper set of complementary modifications is not always straightforward. Modifications can affect other configuration entities, which are also involved in additional constraints that also should not be violated. Thus, the initial changes can propagate throughout the configuration and affect other configuration entities up to the point when all the constraints are satisfied or all the entities have been considered and no solution was found. Such a change propagation process may affect a large number of entities. At runtime this is not desirable because the configuration is a representation of a real system and any change in the configuration has to be applied to the system resources. Thus, the modifications need to be minimal not to destabilize the system. Moreover, to best serve the system requirements the initial configuration is often designed with some optimization in mind resulting in specific values for the different entities/attributes. Finding new values for such entities through change propagation may lead to a consistent configuration but may not be optimal or preserve the configuration designer's preferences. Thus, it is better to keep the change propagation to the minimum and modify the least number of entities.

In this paper we introduce a model-based approach for automating the runtime adjustment of configurations. We extend and generalize the initial approach presented in [23] for the case of single-constraint violation. In general the initial changes are proposed as a bundle (in contrast to a single entity change), and these changes may violate multiple constraints. We define our adjustment approach with respect to the system constraints and the potential impact of the system entities on each other. To limit the number of complementary modifications we determine a propagation scope for each violated constraint. The scope is defined with respect to the impact entities may have on each other, what we call leadership [2]. The leadership concept refines the constraints defined between the configuration entities/attributes. It reflects that some entities/attributes have a dominant or leader role toward others in the constraint. The values of the dominant entities/attributes drive the values of the dominated entities/attributes, which are called followers. Using the leadership concept we can direct and scope the change propagation. Considering a change bundle multiple constraint violations are possible. Moreover, the change propagation for one violated constraint may include an entity, which may also be impacted from the propagation of another violated constraint. We argue that these are related changes and should not be solved independently; in our approach we consider and solve them together. Determining the propagation scopes and identifying the modifiable entities, enable us to formulate the problem as a Constraint Satisfaction Problem (CSP) and use a constraint solver to find the valid set of modifications that solve the constraints. We analyze the complexity of the proposed adjustment approach and conduct experimental evaluations.

The rest of this paper is organized into seven sections. In Section 2 we briefly describe our model-based framework for configuration change management. Section 3 discusses main challenges and introduces the formal definitions of the concepts used throughout the paper. In Section 4 we describe our approach for adjusting a system configuration to maintain its consistency during dynamic reconfiguration. In Section 5 we analyze the complexity of the proposed approach. Section 6 discusses the prototype implementation and the experimental evaluation of the approach. The related work is reviewed in Section 7 before concluding in Section 8.

## 2. MODEL-BASED FRAMEWORK FOR CONFIGURATION CHANGE MANAGEMENT
### 2.1 A Model-based Framework

A system configuration describes the managed resources of a system as well as their relations. The granularity and the definition of the configuration entities depend on the application domain. Components, groups of components, sub-systems, virtual machines, and hardware elements are examples of resources the configuration entities may represent in this paper. The structural and semantic constraints between the configuration entities, their attributes, and their relations are usually defined in a configuration schema. In our work, and following [24], we use UML [16] and its profiling mechanism to define the configuration schema and OCL [17] constraints are added to the configuration profile to capture the domain semantics. These constraints are referred to as consistency rules. A configuration model is consistent if it satisfies all the consistency rules, i.e. all the constraints of its configuration profile.

A constraint is a logical expression that defines a relation between the constrained stereotypes (and their attributes) of the UML profile. A model conforms to a UML profile if and only it respects all the constraints of the profile, thus constraints will be applied on the entities in the model (e.g. classes, nodes, etc.) and their attributes. In the rest of the paper, for the sake of simplicity whenever we talk about constraints between entities we mean the constraints of the profile which are applied on the model entities and their attributes.

In the standard OCL [17] a constraint restricts the possible values of the attributes of a set of entities without giving any preference or role to any of the involved entities. The different entities involved in a constraint are equal and the values of their attributes play equal roles in the satisfaction of the constraint. There is no mechanism to express the fact that in a constraint C involving two entities E1 and E2, one has to choose the values for the attributes of E1 first then accordingly the appropriate values for the attributes of E2 to satisfy the constraint. In other words, there is no mechanism or a construct to express the fact that the value of attributes of E1 force and drive the values of attributes of E2. For instance, if the value of attributes of E1 is changed and the constraint does not hold anymore, we can change the value of attributes of E2 to re-establish the constraint satisfaction, but not the other way around. E2 can change only within the scope valid for E1. We slightly extended standard OCL [2] to express the roles of the entities involved in the constraints as shown in Figure 1. We defined three roles, leader, follower and peer. In an extended OCL constraint we can have some entities in the leader role and the other entities in the follower role. Changes to leader entities may require changes to the follower entities to satisfy the constraint, while changes to follower entities cannot lead to changes to the leader entities even if this is required to satisfy the constraint. The peer role is used in constraints where entities have equal role and they may affect each other mutually.

Note that although these roles are defined for the constrained entities they will actually be applied on their attributes. However, for the sake of simplicity we consider the roles at the entity level instead of attribute level in the rest of the paper.

Figure 2 shows a simplified example of a safe house model in which the constraints (C1, C2 and C3) are shown as ovals linked to the constrained entities with dashed lines labelled with the role of each entity in the constraint. In our example we have three constraints:

C1: If the actual temperature of a room is not equal to the desired temperature, then the air conditioner should be turned on.

C2: If the air conditioner is turned on for a room, then all the windows of the room should be closed and the other way around, if a window in the room is open, the air conditioner is turned off.

C3: If the security level of the security system is set to AllLocked, then all the windows should be closed.

For instance, in constraint C1, the room has the leader role. If constraint C1 is violated because the user has changed the desired temperature or because of the fluctuation of the actual temperature, the air conditioner status has to change to re-establish the satisfaction of the constraint and not the other way around. The status of the air conditioner is driven by the values of the actual temperature and the desired temperature. The air conditioner is a follower entity.

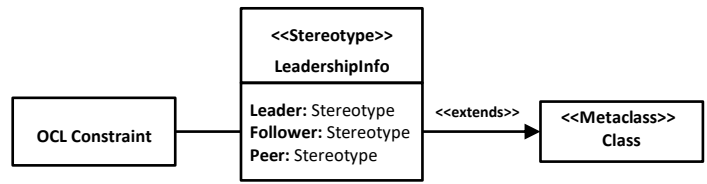

Fig. 1. OCL extended with roles.

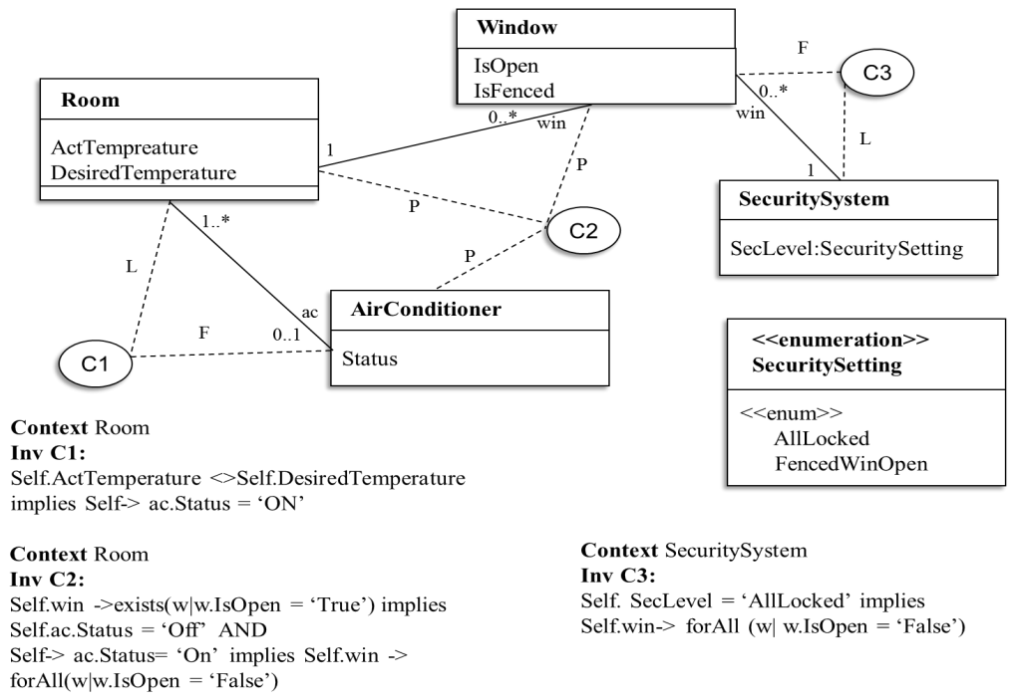

**Context** Room
**Inv C1:**
Self.ActTemperature <>Self.DesiredTemperature
implies Self-> ac.Status = 'ON'

**Context** Room
**Inv C2:**
Self.win ->exists(w|w.IsOpen = 'True') implies
Self.ac.Status = 'Off' AND
Self-> ac.Status= 'On' implies Self.win ->
forAll(w|w.IsOpen = 'False')

**Context** SecuritySystem
**Inv C3:**
Self. SecLevel = 'AllLocked' implies
Self.win-> forAll (w| w.IsOpen = 'False')

Fig. 2. An example of leader/follower/peer entities.

### 2.2 Configuration Change Management

As mentioned earlier, a system reconfiguration may be performed for various reasons, such as in response to environment change or for fine-tuning. Regardless of the reason, the consistency of the configuration should be preserved. We propose the framework shown in Figure 3 for the management of configurations. In this framework a configuration validator is used to check the safety of the requested change (i.e. if the configuration would remain consistent after this change) and an adjustment agent to try to resolve any inconsistency detected in the validation phase. To reduce the validation time, we devised a partial validation approach [2] to identify the constraints relevant to the changes and to be checked, and prune those that do not need to be checked again as they are still valid and not impacted by the changes.

In our partial validation approach, the selected constraints are categorized based on the role of the changed entities involved. Three sets of constraints are defined, LConstraints, FConstraints and PConstraints respectively for constraints where the changed entity is a leader, a follower or a peer. The validation starts with the most restrictive category, i.e. FConstraints. If a constraint in the FConstraints set is violated, the validation is stopped and the changes are rejected. This is because the FConstraints set collects the constraints for the changes of only follower entities and follower entities cannot impact leader entities. On the other hand if a violation is detected in the LConstraints and/or PConstraints we have the possibility to adjust the changed configuration by modifying the respective follower and/or peer entities of the violated constraints, and do this in a recursive manner until all the constraint violations are resolved or no solution is found. This propagation and adjustment are elaborated further in the following sections.

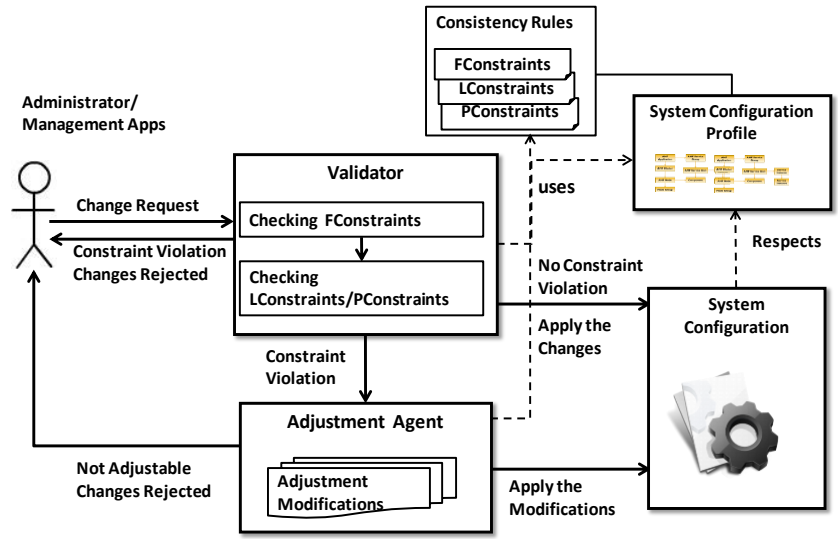

Fig. 3. Configuration change management framework.

### 3. CHALLENGES AND FORMAL DEFINITIONS

The consistency of a configuration should be maintained throughout the system lifecycle to avoid any mal-functioning of the services and applications deploying this configuration. To ensure consistency, any reconfiguration is checked against the consistency rules. Thus, if a constraint is violated because of changes in one or more of its constrained entities, the changes must be rejected unless the other entities involved in the constraint can be modified to satisfy the constraint, i.e. to complement the proposed changes. In turn, these complementary modifications may cause new violations as the modified entities may be subject to other constraints. Thus, these newly violated constraints need to be handled as well. This way the modifications propagate in the configuration model to the point when all the constraints are satisfied i.e. the adjustment is successful, or no further modification is possible while still some constraints are not satisfied, i.e. no successful adjustment is possible and the change has to be rejected.

The adjustment process is defined as a set of complementary modifications and (if necessary) the propagation of these modifications in the model to find a solution which satisfies all the constraints. Thus, a solution includes some entities of the configuration model with new values – requested and adjusted – that along with the other entities of the model satisfy all the constraints. The adjustment process has two steps: first step is to identify the scope of the changes i.e. what entities of the model may need to be modified; and the second step is to modify as few entities as possible in the scope in such a way that all the constraints are satisfied, i.e. address how the modifications should be done. In our approach first we identify the propagation scope for each incomplete change by collecting all the entities that can be affected by the incomplete change. We then try to modify a minimum subset of this scope to satisfy the violated constraints.

In this section we first discuss the challenges we face with the automated adjustment of configuration models, we provide the necessary formal definitions and finally introduce our adjustment approach. We formulate the problem as a CSP and use a constraint solver to find the appropriate complementary modifications.

### 3.1 Main Challenges

**Change Propagation**

The configuration of a large system consists of thousands of interrelated entities. In such models, an attempt to resolve the violation of a single constraint can result in changes of multiple entities which in turn may violate other constraints. The changes may propagate in an exponential manner and finally result in changing a large number of entities (the whole configuration model in the worst case). This is not desirable as more changes results in more constraints to solve which requires more time and computation. Moreover, more changes in the configuration mean the reconfiguration of more system resources in the running system (as the configuration changes need to be applied to system resources to take effect). More reconfigurations in the system in turn risk more the system stability - especially undesirable in highly available systems. Thus, it is desirable to keep the changes to the minimum possible. The issue is how to limit the change propagation to reduce the number of changes and the cost of the changes.

**Multiple Related Changes**

As mentioned earlier, it is possible that multiple changes are requested in a change bundle. Some of these changes may violate some constraint(s) that will trigger change propagation. The propagation of different changes may intersect in the sense that they may affect the same entities. The resolution of such violations may not be possible independently and a solution is only possible when the propagations of multiple changes are considered together, which makes the resolution more complicated. The challenge is how to decide when multiple changes are related and how to consider their propagation together.

### 3.2 Formal Definitions

In our model based framework we define the constraints over the stereotypes of the configuration profile. By applying the stereotypes of the profile to the entities of the configuration models, we ensure that the constraints are also applied to all the instances of those stereotypes. As we have both the configuration model and the profile we can figure out the constraints that are applied to the entities of the model. For the sake of simplicity, we use the constraints as part of the configuration model.

Formally, we define a configuration model as a tuple G = < $En$, C, Role, $f$>, where

- $En$ is a set of configuration entities,

- C is a set of configuration constraints,

- Role is the set of leadership roles for the constrained entities, Role = {leader, follower, peer},

- $f$ is a function defined over the cross-product of entities and constraints, which associates a role with an entity in a constraint

  $f: En \times C \rightarrow Role$, with the following constraints

  o For any constraint if there is a leader entity then there is at least one follower and there is no peer entity (note that we may have more than one leader in a constraint):

  $\forall c \in C: \exists en_x \in En$ with $f(en_x, c) =$ leader $\Rightarrow \exists en_y \in En$ with $f(en_y, c) =$ follower $\land \nexists en_z \in En$ with $f(en_z, c) =$ peer.

  o For any constraint if there is a peer entity all entities involved in the constraint are peer:

  $\forall c \in C: \exists en_x \in En$ with $f(en_x, c) =$ peer $\Rightarrow \forall en_y \in En$ with $f(en_y, c) \neq$ Nil, $f(en_y, c) =$ peer.

We use the term *ChangeBundle* to denote the initial set of changed entities. The subset of entities of the ChangeBundle that causes violation is referred to as *IncompleteChangeSet* and it is obtained from the validation phase. We refer to an entity in the IncompleteChangeSet as an *infringing entity*. An infringing entity is either a leader or a peer entity in the violated constraint because if the entity was a follower, then the change would have been rejected in the validation phase.

IncompleteChangeSet = {$en_x$ ∈ ChangeBundle | ∃c ∈ C with $f$ ($en_x$, c) = (leader or peer) ∧ c is not satisfied}.

The subset of constraints that are violated by the entities of the IncompleteChangeSet is called the *ViolatedConstraintSet* and it is also obtained from the validation phase.

ViolatedConstraintSet = {c ∈ C | ∃$en_x$ ∈ ChangeBundle with $f$ ($en_x$, c) = (leader or peer) ∧ c is not satisfied}

The *SinkSet* of the configuration model contains entities which have only follower or peer roles in all the constraints they are involved in.

SinkSet = {$en_x$ ∈ $En$ | ∀$c_y$ ∈ C with $f$ ($en_x$, $c_y$) ≠ Nil, $f$ ($en_x$, $c_y$) = (follower or peer)}.

We define a binary *Compulsion* relation (▷) among the model entities as follows:

- ∀ $en_i$, $en_j$ ∈ $En$, $en_i$ ▷ $en_j$ ⇔ (∃$c$ ∈ C | $f(en_i, c)$ = leader ∧ $f(en_j, c)$ = follower) ∨ (∃$c$ ∈ C | $f(en_i, c)$ = peer ∧ $f(en_j, c)$ = peer)

or

- ∀ $en_i$, $en_j$ ∈ $En$, $en_i$ ▷ $en_j$ ⇔ ∃ $en_k$ ∈ $En$ | $en_i$ ▷ $en_k$ ∧ $en_k$ ▷ $en_j$

The Compulsion relation is transitive by definition.

**Propagation Scope**

A propagation scope is a slice of a configuration model which contains all the entities and constraints that can be affected through the propagation of a change. The propagation scope for an infringing entity is defined by the set of entities that are in a compulsion relation with the infringing entity and their associated constraints. A violated constraint may have many infringing entities, but their propagation scopes are equal because they are leaders/peers in the same violated constraint and as such they may equally impact all the followers/peers of this violated constraint. Therefore for a violated constraint we consider the propagation scope of one infringing entity only.

For an infringing entity ($en_i$) in a violated constraint ($c_x$) the propagation scope PSi is a tuple < $E_i$, $C_i$, Role, $f_i$> where:

- $E_i$ = {$en_i$} ∪ {$en_j$∈$En$ | $f(en_j, c_x)$=(follower or peer)} ∪ {$en_j$ ∈$En$ | ∃ $en_k$ ∈ $E_i$ with $f(en_k, c_x)$=(follower or peer) ∧ $en_k$ ▷ $en_j$},

- $C_i$ = {c ∈ C | ∃ $en_x$ ∈ $E_i$\{$en_i$} ∧ $f(en_x, c)$ ≠ Nil},

- Role = {leader, follower, peer}, and

- $f_i$ is the project of $f$ on $E_i$ × $C_i$

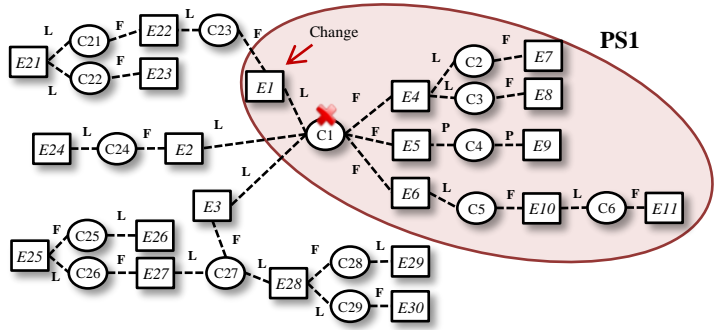

Fig. 4. An example of propagation scope for an infringing entity.

Figure 4 shows an example of a model with an infringing entity E1 which violates constraint C1. In this model the relations shown reflect the entities involvement in the constraints. The propagation scope PS1 for entity E1 is delimited in the figure.

**Propagation Path**

In each propagation scope we can identify a set of propagation paths. A path is defined as an ordered set of entities that starts with an infringing entity and follows the compulsion relation between the entities. It ends if the entity is a sink entity or if it is only leader/peer for entities of the path itself – to avoid cycles. Changes propagate along paths within each scope.

For an infringing entity $en_i$ with the propagation scope PSi, a Path$_x$, is defined as follows:

- $en_i \in \text{Path}_x$,
- $\forall\, en_j \in E_i$, $en_j \in \text{Path}_x$, iff
    - $en_i \triangleright en_j$, and
    - $\forall\, en_k \in \text{Path}_x$, $(en_k \triangleright en_j) \lor (en_j \triangleright en_k)$
- $\exists\, en_k \in \text{Path}_x$ such that
    - $en_k \in \text{SinkSet}$, or
    - $\exists\, en_j \in E_i$, $en_k \triangleright en_j \Longrightarrow en_j \in \text{Path}_x$

The collection of all paths in a propagation scope is called a PathCollection. In addition to the infringing entity, which is common to all the paths in a PathCollection, different paths may have other entities in common as well.

Figure 5 shows the propagation scope for the infringing entity E1 which violates constraint C1. As E1 is a leader entity in C1, its change can propagate to the follower entities of C1 resulting in multiple paths (i.e. Path A, Path B, Path C, Path D). The paths start with E1 and end with an entity with only a follower role or they end with an entity with only a peer role whose other peers have already been visited in the path. E.g. Path A ends with E7 which is a follower role in C2, and Path C ends with E9 which is peer in C4 and its other peer (E5) already exists in the Path C. Note that the propagation scopes and propagation paths can be determined using depth-first search algorithms from the graph theory [20].

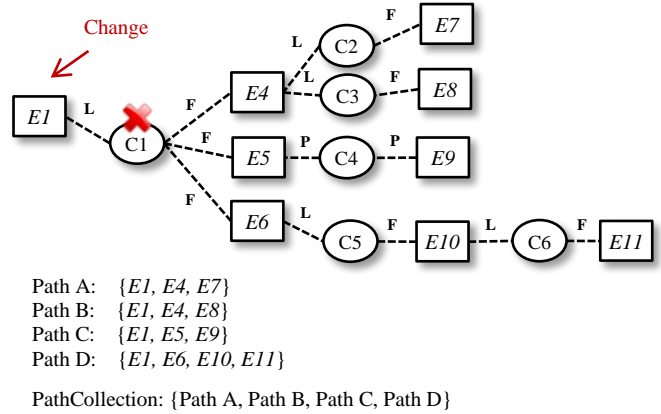

Path A: {*E1, E4, E7*}
Path B: {*E1, E4, E8*}
Path C: {*E1, E5, E9*}
Path D: {*E1, E6, E10, E11*}

PathCollection: {Path A, Path B, Path C, Path D}

Fig. 5. An example of PathCollection with multiple paths for an infringing entity.

## 4. ADJUSTING CONFIGURATION MODELS TO PRESERVE CONSISTENCY

Determining the propagation scope allows us to isolate the problem and ensures that all the entities that can possibly be impacted through change propagation are gathered in the scope. If we find for the propagation scope a set of additional changes with which we can satisfy all the constraints of the scope including the formerly violated constraint cx then we can accept the changed entity of cx as it is not infringing any more. For the example of Figure 5 we can try to find such a solution for E1's violation of C1 by selecting one path at a time until we find such a solution, for example in Path C by adjusting both E5 and E9.

Multiple changes that are requested as a change bundle may cause multiple-constraint violation. In this case we calculate for one of the infringing entities of each violated constraint a propagation scope and PathCollection. If these propagation scopes are disjoint (no common entity between them), then we try to solve each scope independently. On the other hand, if the propagation scopes intersect, we try to solve them together. Our assumption is that a change bundle contains changes, which are bundled together for a reason – they are related. Therefore for such overlapping scopes a solution is more likely when they are considered together as a single problem. Let consider the model in Figure 6 which shows an example of overlapping propagation scopes and their intersection. In this model E1, E12 are two infringing entities violating C1 and C8, respectively. For them the propagation scopes PS1, PS2 and Paths are calculated. Now let us assume that for PS1, Path A is selected and is satisfiable. For PS2 we have only one path, Path X, which is not satisfiable. This results in rejecting the changes of E1 and E12. However, if we select Path B for PS1 then PS2 also becomes satisfiable because the changes in PS1 affect the entities of the intersection of the two scopes. Thus E1 and E12 are acceptable.

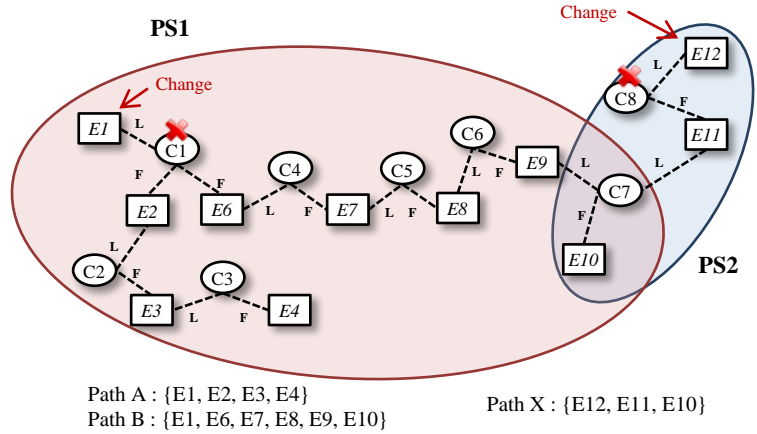

Path A : {E1, E2, E3, E4}
Path B : {E1, E6, E7, E8, E9, E10}
Path X : {E12, E11, E10}

Fig. 6. Multiple constraints violations with overlapping propagation scopes.

### 4.1 Grouping the Overlapping Scopes

Not to miss a solution such as the one discussed above, overlapping scopes need to be solved together. For this purpose we gather overlapping scopes together into a Group. For each Group an Intersect captures the common entities of some overlapping scopes. Different Groups are disjoint (i.e. they have no common entities) as overlapping scopes are collected in the same Group. An example of a Group and its Intersect formed from three propagation scopes PS1, PS2, and PS3 is shown in Figure 7. E30 is the common entity between PS1 and PS2 and E22, E23 are the common between PS2 and PS3. The three scopes form Group1 with the Intersect of E30, E22, and E23.

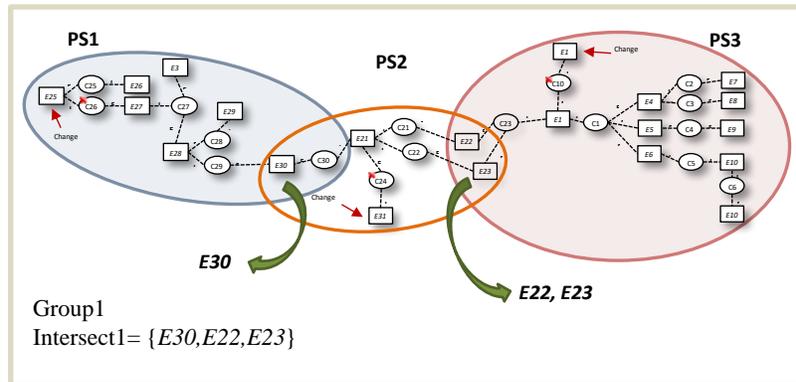

Fig. 7. The formation of a group and its Intersect from overlapping scopes.

For a configuration model the GroupSet is the collection of the Groups and the IntersectSet is the collection of the Intersects of the Groups. To form the GroupSet and its Groups each propagation scope of the model is compared with the already formed Groups of the GroupSet: If the scope has common entities with a Group, it is added to the Group and the Intersect of the Group is updated accordingly with the common entities. If a scope cannot be added to any of the existing Groups, a new Group is created with an empty Intersect. Algorithm 1 describes how the Groups are formed with the overlapping scopes. The input of the algorithm is the collection of all propagation scopes (i.e. PropagationScopeCollection) and the outputs are the GroupSet and IntersectSet. At the beginning both GroupSet and IntersectSet are empty (lines 2-3).

Each propagation scope is compared with the Groups of the GroupSet to find the Groups with which it has common entities (lines 4-20). An integer variable K keeps track of the addition of the scope to a Group, i.e. the index of that Group. At the beginning K is set to -1 (line 5). If a Group is found with entities common with the scope and the scope has not been added to any group yet (i.e. K=-1), it is added to the Group, their common entities are added to the Intersect of the Group, and K is set to the index of the Group (lines 6-11). Since the scope may have common entities with more than one group, it is checked if the scope has common entities with any other groups of the GroupSet. If so, such groups are merged with the first Group to which the scope was added, the Intersect of the Group is updated accordingly and the groups merged into the first one are deleted. (lines 12-18).

If after checking all the Groups no Group was found with common entities (i.e.K is -1), then a new Group is created with the scope and with an empty Intersect (lines 22-25). The grouping procedure is repeated for each scope, and finally the calculated GroupSet and IntersectSet are returned as the output (line 27).

Once the Groups have been created, we try to solve each Group separately. We can distinguish two types of Groups: (1) groups with a single scope (i.e. with an empty Intersect) and (2) groups with multiple scopes (i.e. with a non-empty Intersect). We solve them differently. We use the Depth-first incremental change propagation method for each Group with a single scope, and the Path Bonding method for groups with multiple scopes. Next we discuss these methods.

**ALGORITHM 1.** Grouping the Overlapping Scopes

**Input:** PropagationScopeCollection,

**Output:** GroupSet , IntersectSet

```
1:   // Grouping the overlapping scopes
2:   GroupSet:={}
3:   IntersectSet:={}
4:   For each PSi in PropagationScopeCollection
5:       K:= -1
6:       For (j:=0; j<|GroupSet| ; j++)
7:           If (Epsi∩EGroupj ≠∅) then
8:               If (K == -1)
9:                   Intersectj :=Intersectj ∪ (Epsi ∩ EGroupj)
10:                  Groupj :=Groupj ∪ PSi
11:                  K:=j
12:              Else  ///PSi is already added to Groupk
13:                  Groupk :=Groupk ∪ Groupj
14:                Intersectk:=Intersectk∪Intersectj∪(Epsi∩EGroupj)
15:                  Delete Groupj
16:                  Delete Intersectj
17:                  j--
18:              End if
19:          End if
20:      End For
21:  // If PSi has no intersection with the groups, create a new group with the PSi
     and an empty intersection for that
22:      If (K==-1) then
23:          GroupSet:= GroupSet ∪ {PSi}
24:          IntersectSet:= IntersectSet ∪ { }
25:      End if
26:  End For
27:  Return GroupSet,  IntersectSet
```

### 4.2 Depth-first Incremental Change Propagation for Solving a Group with a Single Scope

The PathCollection of a propagation scope for an infringing entity may contain different paths. To find a solution (i.e. the complementary changes which satisfy the constraints of the scope), one path at a time is selected and tried by modifying the entities in the path. If no solution can be found in the selected path, another path from the PathCollection is selected. The path selection ends when either a solution is found or when all the paths have been exhausted. If all the paths have been exhausted and no solution is found, all the paths of the PathCollection are considered together to find a solution by changing multiple followers/peers of the different constraints. It is possible that no solution exists. However, if there is a solution, we will not miss it by considering all the constraints of the scope if no solution was found in any of the paths individually.

The paths in the PathCollection can be ordered according to their lengths in terms of entities. As we aim at changing the least number of entities we try to find a solution in the shortest path first. Moreover, we use an incremental change propagation to select the minimum number of entities in the path for modifications. In each increment we try to find a new value for a selected follower or peer entity of a violated constraint. In the first increment the selected entity is the second entity of the path which is directly related to the infringing entity of the scope. In the second increment the third entity in the path is selected, which is directly related to the second, and so on. Each selected entity, which is not at the end of the path, participates in two sets of constraints. We refer to the set of constraints, in which the selected entity has the follower role as MandatoryC as it contains the constraints, which must be satisfied by the change of the selected entity. The second set of constraints includes the constraints in which the selected entity has a leader or peer role. We call this set RelaxC. We try to find a change for the selected entity that satisfies the constraints in this set as well, but if we cannot, we relax the problem by dropping these constraints. This is because the selected entity has a leader/peer role in these constraints and if these constraints are violated we can try to resolve them in the next increment by selecting the next entity in the path (which is a follower or peer in RelaxC constraints).

At any increment if a solution is found while considering both sets of constraints the propagation stops. If we cannot find a solution after removing the RelaxC constraints, the selected path is unsolvable. Note that in the last increment when the selected entity is the last entity of the path, we only need to solve MandatoryC as RelaxC is empty.

Algorithm 2 describes our depth-first incremental change propagation process. The inputs of the algorithm are the propagation scope for the infringing entity, the constraint it violates, the PathCollection and the ConstraintSet of the propagation scope. The output is the Solution obtained from the constraint solver. First the paths in the PathCollection are sorted based on their length (line 2). At the beginning MandatoryC consists of the initially ViolatedConstraint. The paths of the PathCollection are selected one at a time and in each selected path we follow the incremental propagation by selecting the next entity and considering its constraints from the ConstraintSet (lines 9-17). If there is a solution that satisfies the constraints in both MadatoryC and RelaxC, then the solution is returned and the algorithm terminates (lines 18-21). Otherwise, if the constraints in MandatoryC are not satisfiable (the Solution is empty), the path is unsolvable (lines 22-23). When the constraints in MandatoryC are satisfiable but the constraints in RelaxC are not, we proceed with the next increment and select next entity of the path and repeat the same procedure until we find a solution, we find that the path is unsolvable, or we reach the end of the path. If the path is unsolvable then all the constraints except the ViolatedConstraint are removed from MandatoryC (line 29) and the next path of the PathCollection is selected. After exploring all the paths if no solution is found, we try to find a solution by considering all paths together, which means giving all the constraints in ConstraintSet of the scope simultaneously to a constraint solver (line 31).

**ALGORITHM 2.** Depth-first Incremental Change Propagation

**Input:** PropagationScope, ViolatedConstraint, PathCollection, ConstraintSet
**Output:** Solution

1: //Sort the PathCollection based on the length of the paths
2: Sort(PathCollection[])
3: UnSolvablePath:= False
4: SolutionFound:=False
5: MandatoryC:={ViolatedConstraint}
6: RelaxC:={}
7: **For** (j:=0; j<|PathCollection|&SolutionFound==False; j++)
8:   SelectedPath:= PathCollection[j]
9:   **For**(i:=1; i<|SelectedPath|&UnSolvablePath==False; i++)
10:     entity:= SelectedPath[i]
11:     **For each** constraint in ConstraintSet
12:      **If** ($f$ (entity,constraint)== leader or peer) **then**
13:       RelaxC:= RelaxC $\cup$ {constraint}
14:      **Else if** ($f$ (entity,constraint)== follower) **then**
15:       MandatoryC:= MandatoryC $\cup$ {constraint}
16:      **End if**
17:     **End for**
18:     Solution:= Solve(MandatoryC $\cup$ RelaxC)
19:     **If** (Solution$\neq${}) **then**
20:      SolutionFound:=True
21:      **Return** Solution
22:     **Else if** (Solve(MandatoryC)=={}) **then**
23:      UnSolvablePath:= True
24:     **Else**
25:      MandatoryC:=MandatoryC $\cup$ RelaxC
26:      RelaxC:={}
27:     **End if**
28:   **End for**
29:   MandatoryC:={ViolatedConstraint}
30: **End for**
31: Solution:= Solve(ConstraintSet)
32: **Return** Solution

Figure 8 shows an example application of the depth-first incremental propagation. In this figure E1 is the infringing entity and C1 is the ViolatedConstraint. At first as shown in stage (a) of the figure, Path A which is one of the paths with the shortest length in the PathCollection is selected (Path B and Path C also have the same length and could have been selected). In Increment 1 shown as stage (b) in the figure, the second entity of Path A, i.e. E4, is selected for the change propagation. C1 is added to MandatoryC and because E4 has the leader role in C2 and C3, these constraints are added to RelaxC. As we cannot find a new value for E4 to satisfy the constraints in both MandatoryC and RelaxC, the problem is relaxed by disregarding RelaxC constraints. Assuming that a new value can be found for E4 that satisfies the MandatoryC constraints therefore RelaxC is added to MandatoryC, RelaxC is emptied and we proceed to Increment 2 which is shown in stage (c) of Figure 8. In this increment the next entity of Path A, i.e. E7, is selected. E7 has a follower role in both C2, C6 (i.e. E7 is a sink entity), thus both constraints are added to Mandatory C and because E7 is the last entity of the Path, RelaxC remains empty. If new values can be found for E7 and E4 that satisfy all the constraints of MandatoryC (i.e. C1, C2, C3, C6), we have a solution, otherwise another path should be explored.

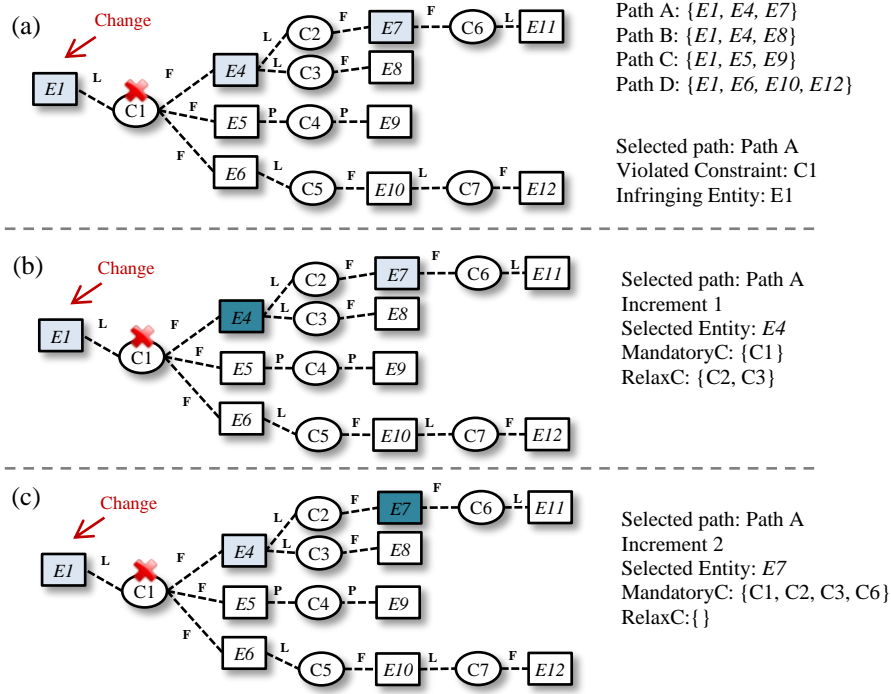

Fig. 8. Application of depth-first incremental change propagation.

### 4.3 Path Bonding for Solving Groups with Multiple Scopes

The scopes of a Group need to be solved together. To avoid changing all entities of a Group and reduce the number of changed entities, we proceed as follows: In each Group all the paths which have entities common with the Intersect of the Group are selected to form the BondedPath of the Group. In other words we bond the related paths and disregard the other paths which do not have common entities with the Intersect of the Group. The entities of the bonded paths of each Group are our primary candidates for the complementary changes.

Figure 9 shows the path selection for Group1. For the sake of simplicity only the first and the last entities of each path are shown explicitly while other entities of the paths are represented by the arrow between the first and the last entities of the path. For path bonding we select all the paths which have at least an entity in the Intersect of the Group.

By grouping the scopes and bonding their paths we address the fact that the initial changes requested in the same change bundle are related to each other. Thus, when potential inconsistencies are detected during the validation of a requested change bundle, it is most likely that the solution is possible only by considering the related scopes together.

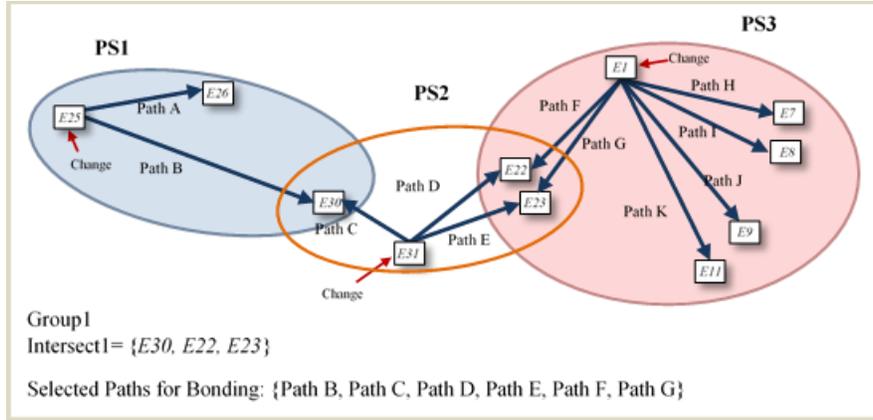

Fig. 9. Selecting the paths of the Group for the bonding.

Similarly to the depth-first incremental change propagation, for each entity in the BondedPath, the mandatory constraints (MandatoryC) need to be identified and satisfied by the complementary changes. The MandatoryC in this case contains all the constraints in which the entities of the BondedPath are participating (as leader, follower or peer). Unlike in case of single-scope groups , we cannot perform an incremental propagation, thus, there is no need for the RelaxC. We try to find a solution by considering the BondedPath of a Group. If we cannot find a solution that satisfies all the constraints in MandatoryC, then we need to consider the other paths of the Group as well. Algorithm 3 describes the path bonding and finding the solution for groups with multiple scopes. The Group, its PathCollectionSet, its Intersect, its ConstraintSet and the IncompleteChangeSet are provided as input to the algorithm and the output is the Solution for the group. For the given Group a BondedPath, a MandatoryC and a Solution are initialized first (lines 1-3). The BondedPath is calculated for the Group, i.e. for each infringing entity of the IncompleteChangeSet that also belongs to the Group (line 5). We select the paths from thePathCollectionSet that have common entities with the Intersect of the Group (line 6). The identified paths are added to the BondedPath of the Group (lines 6-10). Next the constraints related to the entities of the BondedPath are added to MandatoryC (lines 13-19). A solution for the Group should satisfy all the MandatoryC constraints (line 20). If such a solution exists (Solution is not empty), it is returned as the output, i.e. the solution for the Group (lines 21-22). If the BondedPath is not satisfiable (Solution is empty) then all the constraints of the Group are considered and are added to MandatoryC (lines 23-31). The result of solving the MandatoryC is the solution of the Group (line 32). Finally, the Solution (empty or not) is returned as the output of the algorithm (line 34).

**ALGORITHM 3.** Bonding the Related Paths and Their Resolution

**Input:** PathCollectionSet, Group, Intersect, IncompleteChangeSet, ConstraintSet

**Output:** Solution

1: Solution:={}
2: BondedPath:={}
3: MandatoryC:={}
4: // Bonding the related paths of each Group
5:   **For each** $en_j$ in IncompleteChangeSet && $Group_i$
6:     **For each** $path_k$ in $PathCollection_j$
7:       **If** ($path_k \cap Intersect_i \neq \{\}$) **then**
8:         BondedPath:=BondedPath $\cup$ $path_k$
9:       **End if**
10:    **End For**
11:  **End for**
12: //Identifying the constraints related to the entities of the bonded path
13:  **For each** $en_x$ in BondedPath
14:    **For each** $c_y$ in ConstraintSet
15:      **If** ($f(en_x, c_y) \neq Nil$) **then**
16:        MandatoryC:= MandatoryC $\cup$ $\{c_y\}$
17:      **End if**
18:    **End for**
19:  **End for**
20:  Solution:=Solve(MandatoryC)
21:  **If** (Solution$\neq$\{\}) **then**
22:    **Return** Solution
23:  **Else**
24:    MandatoryC:={}
25:    **For each** $en_x$ in $E_{group_i}$
26:      **For each** $c_y$ in ConstraintSet
27:        **If** ( $f(en_x, c_y) \neq Nil$) **then**
28:          MandatoryC:= MandatoryC $\cup$ $\{c_y\}$
29:        **End if**
30:      **End for**
31:    **End for**
32:    Solution:=Solve(MandatoryC)
33:  **End if**
34: **Return** Solution

### 4.4 Selecting the Method for Solving the Groups

Algorithm 4 describes the overall approach for the adjustment, which includes the two methods for solving the groups. The depth-first incremental propagation method is used for the groups with a single scope and path bonding for the groups with multiple scopes.

The input of the algorithm consists of the PathCollectionSet, GroupSet, IntersectSet, IncompleteChangeSet, ConstraintSet, and the ViolatedConstraintSet. The algorithm tries to solve each Group in the GroupSet resulting in a PartialSolution. The final Solution for the incomplete change set is the union of all these PartialSolutions or it is empty. The output is empty if any of the Groups is unsolvable; otherwise it contains the complementary changes for adjusting the configuration model for the changes in the IncompleteChangeSet.

The algorithm starts by initializing the Solution as an empty set (line 1). For each Group in the GroupSet a PartialSolution is initialized (line 1). If the Intersect of the Group is Null, i.e. there is only one scope in the Group then the depth-first incremental change propagation method is used for solving the Group (lines 5-8). If the Intersect of the Group is not Null, i.e. there is more than one scope in the Group, then the path bonding method is used to solve the Group (lines 9-12). In either case the PartialSolution of the Group is obtained. If the PartialSolution is not empty (i.e. the Group is solvable), the returned PartialSolution is added to the final Solution (lines 13-14) and the procedure repeats for the next Group. If a PartialSolution is empty, which means a Group is not solvable and thus Solution is also emptied (lines 15-18) as there is no solution. The adjustment is not possible; the inconsistencies caused by the infringing entity/entities cannot be resolved. At the end if all Groups in the GroupSet are solvable, the ultimate Solution is returned (line 20) containing all the complementary changes needed for the adjustment.

## 5. COMPLEXITY ANALYSIS

In our solution we use some heuristics to reduce the number of complementary modifications. However, they introduce some overhead. As the overall approach

---

**ALGORITHM 4. Overall Approach for Adjustment**

**Input:** PathCollectionSet, GroupSet, IntersectSet, IncompleteChangeSet,
  ConstraintSet, ViolatedConstraintSet

**Output:** Solution

1: Solution:={}
2: **For each** $Group_i$ in GroupSet
3:   PartialSolution:={}
4:   ViolatedConstraint:={}
5:   // Incremental Propagation is called for a group with a single scope (Null Intersect)
6:   **If** ( $Intersect_i$==Null) **then**
7:     ViolatedConstraint:=Select($scope_j$, ViolatedConstraintSet)
8:     PartialSolution:= Depth-firstIncrementalPropagation (Propagation $Scope_j$, ViolatedConstraint, $PathCollection_j$, $ConstraintSet_i$)
9:   // BondingPath is called for a group with multiple scopes
10:  **Else**
11:    PartialSolution:=BondingPath(PathCollectionSet, $Group_i$, $Intersect_i$, IncompleteChangeSet, $ConstraintSet_i$)
12:  **End if**
13:  **If** (PartialSolution ≠{})
14:    Solution:=Solution ∪ PartialSolution
15:  **Else**
16:    Solution:={}
17:    **Return** Solution
18:  **End if**
19: **End for**
20: **Return** Solution

---

consists of different algorithms, we discuss the time required for each algorithm separately. The time for the overall approach is obtained by the sum of the execution times of these different algorithms.

The adjustment resolution starts by determining the propagation scope and PathCollection for the infringing entities. The propagation scope and the PathCollection can be calculated simultaneously as they follow the same logic. The complexity is the same as for traversing a graph with depth-first search and in worst case it is $O(b^d)$ where $b$ is the branching factor (in our case the number of constrained entities in the constraints) and d is the depth of the search (in our case the longest path length) [20]. If there are no cycles in the configuration model between the constrained entities (i.e. it is a tree), the complexity of the scope/path creation is $O(n)$ where n is number of constrained entities in the model; in the worst case all entities are visited once. This calculation is done m times where m is the number of violated constraints. Thus, the execution time is $m \times (b^d)$ in the worst case and when there is no cycle (tree-based structure) it is $m \times n$.

The second part is the grouping of overlapping scopes. In Algorithm 1 every scope is checked with the existing groups for common entities. So, the worst case scenario is when we have a maximum number of groups. The maximum number of groups is equal to the number of scopes (when each group has only one scope) and in worst case scenario the number of scopes is equal to the number of violated constraints (i.e. infringing entities are violating different constraints and make distinct scopes). For the first scope the algorithm checks 0 groups and creates the first group, the second scope is checked with one group and creates the second group. This continues until it reaches to the m$_{th}$ scope. For the m$_{th}$ scope it checks $(m-1)$ group. So, in total the algorithm performs $(m-1) + (m-2) + (m-3) + \cdots + 1 + 0$ checks and therefore the execution time of the grouping algorithm is $m^2$.

The third part of the adjustment approach is the solving of the groups; using the depth-first incremental propagation for groups with a single scope and the path bonding for the groups with multiple scopes. The execution time of the depth-first incremental propagation for each scope includes sorting the PathCollection and traversing the paths. If $\alpha$ is the number of groups with a single scope, the execution time for solving all groups with a single scope is $\alpha \times p \times log(p) + \alpha \times p \times d$, where $p$ is the average number of paths in the groups, and $d$ is the average path length.

On the other hand for solving each group with multiple scope by path bonding, all paths of each scope are checked to select the ones which have common entities with the intersect of its group. Therefore, the time for solving all groups with multiple scopes is $(m - \alpha) \times p \times d$, where $p$ is the average number of paths in the groups, and $d$ is the average path length and m is the number of scopes (i.e. the number of violated constraints).

Thus, the execution time of the overall adjustment approach is as follows:

Execution time (Propagation scope) + Execution time (Grouping) + Execution time (Depth-first incremental propagation /BondedPath), which is:

$m \times b^d + m^2 + \alpha \times p \times log(p) + \alpha \times p \times d + (m - \alpha) \times p \times d$, where

- $n = $ Total number of entities in the model
- $m = $ Number of violated constraints $= $ Number of scopes
- $b = $ Branching factor (average number of constrained entities in constraints)
- $d = $ The average path length
- $p = $ Average number of paths in the scope

- $\alpha = Number\ of\ groups\ with\ a\ single\ scope\ (i.e.\ a\ portion\ of\ m)$

The complexity is therefore the same as for traversing a graph with depth first search, i.e. $O(b^d)$. Although the determination of the scopes and the usage of heuristics may seem to impose some overhead, the approach has several advantages compared to the traditional constraint solving solutions:

1) We try to limit the scope of the problem by calculating the propagation scopes and using the discussed adjustment heuristics which reduce the complexity of the problem. The complexity of SMT problems is exponential or even worse when it comes to the combination of different theories (e.g. linear integer arithmetic, theory of arrays, etc.) [21, 22].

2) In general constraint solving solutions aim at finding valid values for all the variables of all the constraints and do not consider minimizing the number of changes as we do for the adjustment of configuration models at runtime.

## 6. PROTOTYPE IMPLEMENTATION AND EXPERIMENTAL EVALUATION

We implemented a prototype of our adjustment approach in Microsoft visual studio and used Microsoft Z3 [18] as our constraint solver. To evaluate the performance we used an ETF (Entity Types File) model [19] as our configuration model, which is a component catalog in the availability domain. The UML profile of the ETF model consists of 26 stereotypes and 28 constraints. The initial model conforming to this profile contains 40 entities with a total number of 85 attributes.

To translate this model to a constraint satisfaction problem we follow the partial evaluation of constraints proposed by Song et al. in [9]. A variable is created for each entity or attribute of the model which is involved in a constraint. During the translation process the constraints are also partially evaluated and constraint instances are generated with the variables. For each constraint in the profile we may generate multiple constraint instances (depending on the number of model entities conforming to the context stereotype of the constraint). The created variables and constraint instances are the input of our prototype together with the incomplete change set (subset of the Change Bundle) and the violated constraints (obtained from the validation phase). The tests are performed on a machine with an Intel® Core™ i7 with 2.7 GHz, 8 Gigabytes RAM and Windows 7 as operating system.

### 6.1 Evaluation Scenarios

We measure the execution time and number of necessary complementary changes with our approach, and compare them to the measurements for runs for the "total change" resolution approach where the changes are given to the solver with the subset of the configuration model that they can impact directly or indirectly (i.e. without considering any leadership information). We consider our adjustment resolution in two cases: in the first case, i.e. the overall adjustment resolution, we use the created paths and solve each group with the paths (i.e. use of depth-first incremental propagation for the groups with one scope and use of path bonding for the groups with multiple scopes in the group) and in the second case, which we call it the "group-based" resolution, the complete set of calculated groups of the propagation scopes are given to the solver. We should indicate that the second case (group-based resolution) is actually the worst case of the overall adjustment; this means that if no solution can be found by depth-first incremental propagation or path bonding for a group, then the whole group is considered for the modifications.

Three scenarios are considered: (1) solving a group with a single scope, (2) solving multiple groups, and (3) detecting the not-adjustable changes for multiple groups. For the first two scenarios we measure the number of complementary changes and the time needed for calculating them. For the last scenario we only do the comparison of the execution time of the overall adjustment resolution with the total change resolution.

### 6.2 Solving a Group with a Single Scope

In this scenario we consider 10 test cases. In each test case one entity is changed randomly to violate a constraint. For each test case we measure the execution time and the number of necessary complementary changes of the adjustment resolution (more specifically the depth-first incremental propagation method) and compare them to the execution time and number of complementary changes for the group-based resolution and for the total change resolution. The results are shown in the diagrams of Figure 10 and Figure 11. The results reported in Figure 10 show that the number of complementary changes with the overall adjustment resolution is always less than the total change resolution and it is also less or in some test cases equal to the number of changes for group-based resolution. The number of changes are equal for the two resolutions (overall adjustment and group-based) when the depth-first incremental propagation is unable to solve the scope by considering a single path and has to consider all the paths of the scope together or when there is only a single path in the scope and the number of increments is equal to the path length (both cases make the incremental and group-based resolution appear the same).

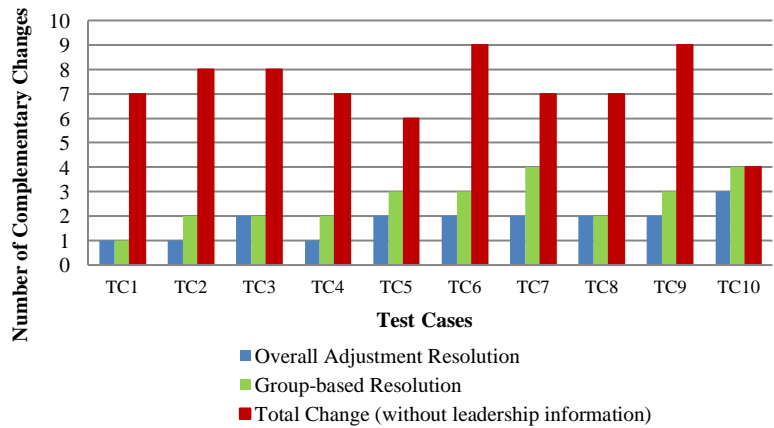

Fig. 10. Comparison of the number of complementary changes using the overall adjustment resolution versus the group-based and total change resolution.

Figure 11 shows the comparison of the execution times of the overall adjustment resolution with the execution times of the group-based resolution and measurements of the total change approach. The measured times indicate that the overall adjustment resolution (i.e. the depth-first incremental propagation in this scenario) usually takes less time than the group-based and total change resolutions. However, in test case TC10 the execution time of the overall adjustment is higher than the execution time of the group-based resolution. The reason is that in this test case the depth-first incremental propagation was unable to find a solution in one path and had to consider the group (containing one scope) for the change (similar to the group-based resolution) but as it has already tried each path individually, the total time for the overall adjustment is added and becomes higher.

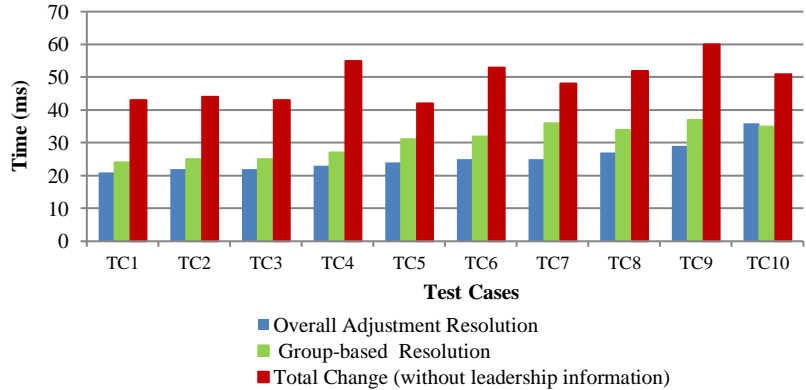

Fig. 11. Comparison of the execution time using the overall adjustment resolution versus the group-based and total change approaches.

### 6.3 Solving Multiple Groups

In this scenario we consider 14 test cases each with a random IncompleteChangeSet. Similarly to the previous scenario we measure the execution times and the number of necessary complementary changes when using our adjustment and compare them to the measurements of group-based and total change resolutions. Figure 12 and Figure 13 show the results of our experiments. We should indicate that measurements for the overall adjustment resolution includes the group creation and solving the groups one by one with either depth-first incremental propagation or by path bonding and, if no solution was found we try to solve by considering the whole group (similar to the group-based resolution).

As the chart in Figure 12 shows, the number of complementary changes of the overall adjustment approach is always less than the number of changes for the group-based or total change resolutions. In our tests it happened that one or two groups would require the resolution by group-based resolution but as the other groups could be solved by the depth-first incremental propagation or by path bonding, the overall adjustment resolution has a better overall outcome and reduces the number of complementary changes in each test case.

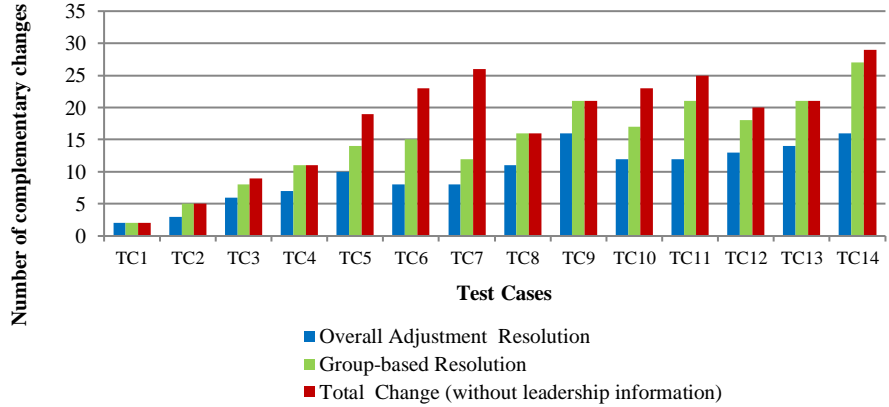

Fig. 12. Comparison of the number of complementary changes using the overall adjustment resolution versus, the group-based and the total change resolutions.

The comparison of the execution times of the overall adjustment, group-based and the total change resolutions shown in Figure 13 also indicates that our resolution is faster than the total change in most of the cases. The test cases in which our resolution approach has a similar or higher execution times compared to the total change resolution are the situations that at least one group could not be solved with the depth-first incremental propagation or the path bonding, thus the whole group is considered to be changed similarly to the group-based resolution.

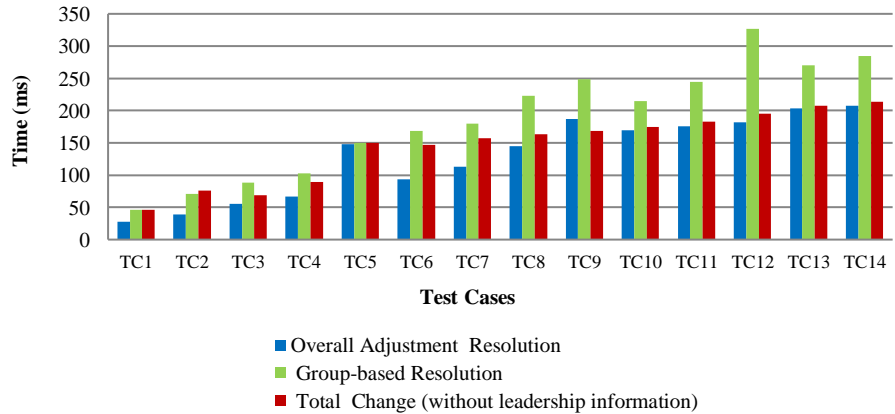

Fig. 13. Comparison of the execution times using the overall adjustment resolution versus the group-based and the total change resolutions.

Overall consideration of the measurements and the comparisons show that our adjustment resolution reduces the number of complementary changes but this is achieved by doing some further calculations (for creating the paths and groups) that can be time consuming, especially when the constraint violations increases.

### 6.4 Finding Not-adjustable Changes for Multiple Groups

Another scenario which we considered for evaluating our proposed approach is to find out how fast it can detect the not-adjustable changes. For this scenario we compare the execution times of our overall adjustment resolution with the execution time of the total change resolution approach and disregard the group-based resolution approach. The reason is that in this scenario the overall adjustment includes the group-based resolution and if no solution can be found with the paths, the whole group is considered for change. Six test cases have been considered, each with different number of constraint violations. Figure 14 shows the execution times for the overall adjustment approach and the total change resolution approach. As the measurements indicate our approach can detect the unsolvable cases faster and this is because each group is tried for resolution independently from the other groups and if a group is unsolvable we can stop the process (because a solution is complete only when all the groups are solved). On the other hand the total change resolution approach considers all the changes and their respective constraints all together which requires more execution time.

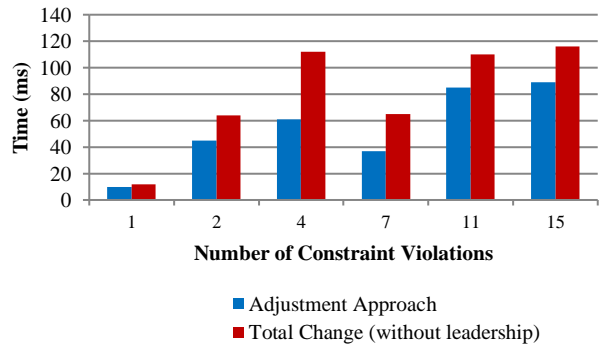

Fig. 14. Comparison of the execution times of the overall adjustment resolution versus the total change resolution.

The experiments indicate that despite the overhead of the propagation scope/path calculations, our approach still outperforms the total change resolution approach in most of the cases by reducing the execution time, reducing the number of complementary changes and by detecting the unsolvable cases faster. However, the execution time in some cases is increased if the depth-first incremental propagation or the path bonding methods cannot find a solution and the whole group is required to be considered for change.

### 7. RELATED WORK

Constraint solving is widely used for configuration generation and adaptation [3, 4, 5, 6]. In [6] the authors proposed a range fix approach that is based on constraint solving. Instead of finding a specific value for a configuration entity/attribute they find ranges (options) that fix the violated constraints. Although the ranges give the user more options to choose, but still requires the user to have knowledge about the configuration so he/she can select values from the ranges. In our approach we automate the configuration adjustment to decrease the user's involvement to reduce the complexity and the risk of inconsistency. We also try to minimize the adjustments not to destabilize the system at runtime. Authors in [4] use constraint solving to automate the configuration generation in Software Product Line (SPL). They follow a multi-step approach using the feature model and a set of constraints for selecting the features (constraints such as cost or priority). At the end of each step a valid configuration with a subset of features is created and the desired target configuration is obtained in the last step. The configurations are created offline, and large modifications may be applied in each step. [3] combines goal modeling with constraint solving for creating configurations for SPL that meet QoS requirements. Authors claim the result is also useful for runtime adaptation. [5] also proposes a constraint guided adaptation framework that formalizes the non-functional requirements of the system as constraints. A symbolic constraint satisfaction method based on Ordered Binary Decision Diagram is used to find a solution. Compared to our approach we handle consistency of the reconfigurations at runtime, thus we are concerned about minimizing the modifications.

Taking into account user preferences in self adaptive systems is discussed in [7, 8, 9]. [8] uses a utility function to formulate the user preferences as an optimization problem for dynamic configuration of resource-aware services. In [9] the user preferences are considered to adapt the runtime models. Their objective is to solve the CSP by satisfying as many constraints as possible. After diagnosing the interrelated constraints, less important constraints (with lower weight) are ignored to satisfy the remaining constraints. Users can revise the model or modify the weight of constraints in order to express their preferences. In our work constraints cannot be ignored, however our approach directs the adaptation (propagation) to relax the problem for interrelated constraints. Our adjustment approach aims at reducing the role of the user in the process. Instead, the role (impact) of the entities in relation to each other is a key feature in our adjustment approach.

Model repair [11] has been a very active research topic for the model driven engineering community. Model repair is about fixing inconsistencies that may arise during the model building/refinement activities. These inconsistencies may arise because of multiple collaborating teams manipulating the model concurrently or simply during the refinement of a model by one designer. For an overview, an up-to-date classification of existing techniques is provided in [[11]. Recently, [12] proposed an approach to repair inconsistencies caused by edit operations during the activity of model design. The inconsistencies are repaired with complementary edit operations. The work is focusing on inconsistencies between diagrams such as, for instance, a method call is added in a sequence diagram but not yet defined in the class diagram. The approach does not look into the constraint satisfaction. Closer to our work, [13] proposed an approach to identify a set of valid choices (values) for each model entity/attribute through incremental consistency checking. The authors argue that the result is a set of choices which fix the initial inconsistency while it does not violate any other constraint. However, it is not always possible to find the set of valid values or the set includes numerous members (e.g. for the attributes with integer or string datatypes). Moreover, the work does not handle interrelated constraints. [14] designed a repair semantics which maps the constraints to repair actions. This approach does handle interrelated constraints and the repair action for one constraint may violate another constraint. [15] defined a language similar to OCL for describing the constraints and also to define the fixes in case of violation of each constraint. However, for defining the fixes, the developer has to consider all the relations between the constraints and reflect them in the fixes. In addition, analyzing numerous invalid values of the model entities requires defining numerous fixes. These challenges make the development of the fixes very complicated especially for a large number of constraints.

## 8. CONCLUSION

To assure the consistency of system configurations during dynamic reconfigurations, the changes need to be checked. The consistency of a configuration has to be preserved at runtime as inconsistency means incorrect data which may result in the malfunctioning of the system. At runtime, inconsistencies in a configuration often happen because of incomplete reconfiguration changes that need to be completed with complementary modifications, i.e. configuration adjustments. Configuration adjustment requires a comprehensive knowledge of the configuration entities/attributes, their relations and system constraints. Not all users (admin or a management application that requests the reconfiguration) have such knowledge or the information may not be exposed to the user, e.g. for security concerns. Moreover, the complementary modifications of the configuration should be kept to the minimum to reduce the time and computational cost of changes and not to destabilize the system at runtime.

We proposed a model-based approach for the adjustment of configurations to address the aforementioned challenges. The structure of the configuration – including the entities and their relations – is captured in a configuration profile. The constraints of the configuration are expressed through extended OCL constraints, which also capture the roles of the configuration entities in the constraints. The leader/follower/peer roles define which entity can impact and drive the other ones in the constraint. At runtime a validator detects potentially incomplete changes that violate configuration constraints. The result of the validation is provided as input for the adjustment engine. In the proposed adjustment approach, a propagation scope is identified with respect to any of the infringing entities. This scope consists of the entities, and related constraints, that may be affected through change propagation. Different change propagation paths are defined within the scope based on the impact of the entities on each other. If the propagation scopes were disjoint, each scope can be handled independently following the method of incremental propagation for the shortest path of the scope. This let us reduce the side-effects of the change propagation and avoid changing entities unnecessarily.

The overlapping propagation scopes need to be solved together assuming that the reconfiguration changes are introduced as a bundle because of the relation between the changed entities. Thus, we try to relate them by bonding the change propagation paths. For each infringing changed entity we select the path which has entities in common with other scopes, and the paths of the overlapping scopes are considered as a single problem to be solved. The defined problem is then given to a constraint solver to determine the new values that satisfy the constraints. As we formulate the adjustment approach by determining the scope and selecting the entities that are required to be modified, we attempt to minimize the modifications in the configuration.

We analyzed the complexity and performed some experimental evaluation of our approach, which demonstrate its efficiency, from the execution time and the number of complementary changes required, compared to other approaches. As next step we plan to investigate the optimization of the scope calculation and other heuristics for the change propagation.

**ACKNOWLEDGMENT**

This work has been partially supported by Natural Sciences and Engineering Research Council of Canada (NSERC) and Ericsson.